\documentclass[graybox]{svmult}

\usepackage{mathptmx}       
\usepackage{helvet}         
\usepackage{courier}        
\usepackage{type1cm}        
\usepackage{makeidx}         
\usepackage{graphicx}        
\usepackage{multicol}        
\usepackage[bottom]{footmisc}
\usepackage{amsfonts}
\usepackage{amsmath}
\usepackage[
      colorlinks=true,
      linkcolor=blue,
      urlcolor=magenta,
      filecolor=green,
      citecolor=red,
      pdfstartview=FitV,
      pdftitle={},
        pdfauthor={Daniel Grumiller and Wout Merbis},
        pdfsubject={},
        pdfkeywords={},
        pdfpagemode=None,
        bookmarksopen=true
      ]{hyperref}
\usepackage{color}

\makeindex             

\begin{document}

\title*{Free energy of topologically massive gravity and flat space holography}
\author{Daniel Grumiller and Wout Merbis}
\institute{Daniel Grumiller and Wout Merbis \at Institute for Theoretical Physics, TU Wien, Wiedner Hauptstrasse 8-10, A-1040 Vienna, Austria, \email{grumil@hep.itp.tuwien.ac.at, merbis@hep.itp.tuwien.ac.at}}
%
%
\maketitle

\vspace{-8cm}
\hfill TUW-15-18

\vspace{8cm}
\abstract{%
We calculate the free energy from the on-shell action for topologically massive gravity with negative and vanishing cosmological constant, thereby providing a first principles derivation of the free energy of Ba\~nados-Teitelboim-Zanelli (BTZ) black holes and flat space cosmologies. We summarize related recent checks of flat space holography.
}

\section{Introduction}
The Schwarzschild solution was found a few weeks after Einstein's theory of general relativity was finished and has engendered a century worth of interesting research results. One indirect outcome of Schwarzschild's remarkable discovery is black hole holography, which is at the core of numerous current research avenues, not just in classical and quantum gravity, but even in neighbouring fields such as quantum field theories or condensed matter physics (specifically at strong coupling).
The anti-de~Sitter/conformal field theory (AdS/CFT) correspondence \cite{Maldacena:1997re} provides a concrete realization of holography \cite{Hooft:1993gx}. 
A specific set of applications and checks is the determination of correlation functions on the gravity side \cite{Gubser:1998bc}
. The 0-point function, or on-shell action, gives the free energy and should therefore capture all features of the free energy of the dual CFT \cite{Witten:1998zw}. The 1-point functions, or vacuum expectation values, allow to determine conserved charges like mass or angular momentum \cite{Balasubramanian:1999re,Henningson:1998gx,deHaro:2000xn}. The 2- and 3-point functions are highly constrained by symmetries and allow basic checks of the correspondence, while the higher $n$-point functions provide further applications and consistency checks, see \cite{Aharony:1999ti} and references therein.

If holography is a true aspect of Nature it must also work beyond AdS/CFT, particularly in flat space. After some early progress on extracting features of the S-matrix as a limit from AdS/CFT correlators \cite{Polchinski:1999ry,Susskind:1998vk,Giddings:1999jq} it took a while to come up with the first precise proposal for a holographic correspondence between a specific quantum theory of gravity in flat space, dubbed flat space chiral gravity \cite{Bagchi:2012yk}, and a specific quantum field theory. This proposal is based on a scaling limit of topologically massive gravity (TMG) \cite{Deser:1982vy}, which is a three-dimensional gravity theory that consists of the Einstein--Hilbert action and a gravitational Chern--Simons term. In the limit of interest only the gravitational Chern--Simons term remains and the ensuing theory, known as conformal Chern--Simons gravity (CSG), has interesting holographic properties \cite{Afshar:2011qw,Bagchi:2012yk,Bertin:2012qw}.

One particular check of the flat space chiral gravity proposal and more generally of flat space holography is the microscopic derivation of the entropy  \cite{Barnich:2012xq,Bagchi:2012xr} of flat space cosmology solutions using properties of the dual Galilean CFT \cite{Bagchi:2010zz}. For flat space Einstein gravity the microscopic result matches the expected Bekenstein--Hawking law, which can be derived from first principles using the Euclidean path integral formulation \cite{Detournay:2014fva} to determine the free energy and extract from it the entropy using standard thermodynamical relations.

Naively applying the same methods to TMG, including its limiting case CSG, appears to fail. Indeed, inserting for instance the BTZ line-element into the bulk-plus-boundary action constructed in \cite{Guica:2010sw} and used in \cite{Afshar:2011qw} yields the correct 1-, 2- and 3-point functions, but not the correct 0-point function or free energy. One can avoid this issue by directly calculating the entropy \cite{Kraus:2005zm}, e.g.~using Solodukhin's method of conical deficits \cite{Solodukhin:2005ah} or Tachikawa's generalization of the Wald entropy \cite{Tachikawa:2006sz}. If one then postulates the first law and integrates it (essentially Legendre transforming entropy) one can extract free energy. However, this indirect derivation of free energy is not completely satisfactory since it uses the first law as an input rather than providing it as a result.

It is of interest to directly calculate free energy for TMG from first principles, since this provides a check of the validity of the first law. In this work we achieve this goal for TMG and its CSG limit, both in AdS and in flat space. 

This proceedings contribution is organized as follows.
In section \ref{se:1} we display the bulk action in a Chern--Simons like formulation and clarify our notation.
In section \ref{se:2} we review the Euclidean BTZ solution.
In section \ref{se:3} we determine the BTZ free energy from the on-shell action.
In section \ref{se:4} we apply our results to flat space cosmology solutions.
In section \ref{se:5} we comment on correlation functions in flat space holography. 
In section \ref{se:6} we conclude with mentioning a further possible check of the flat space chiral gravity proposal, namely holographic entanglement entropy.

\section{Action and notation}\label{se:1}

It turns out to be useful to employ the Chern--Simons-like formulation of TMG, whose bulk action is given by (see \cite{Merbis:2014vja} and references therein)
\begin{align}
I_{\scriptscriptstyle \mathrm{TMG}} =  \frac{1}{4\pi G} \int 
{\rm tr} \Big[  - \sigma e \wedge R + \tfrac{\Lambda_0}{3} e \wedge e \wedge e + f \wedge T - \tfrac{1}{2\mu} \omega \wedge \big( d \omega + \tfrac23 \omega \wedge \omega \big) \Big]  \,.\label{STMG} 
\end{align}
Here $e$ is the dreibein, $\omega$ the (dualized) spin-connection and $f$ is an auxiliary $so(2,1)$ valued one-form field. $R$ denotes the Riemann curvature two-form and $T$ is the torsion two-form. In addition, $\sigma =\pm1$ is a sign parameter and the cosmological parameter $\Lambda_0$ is related to the cosmological constant as $\Lambda_0 = \sigma \Lambda$. In this work we assume that the cosmological constant is either negative (AdS) or vanishes (flat space). The quantity $G$ is Newton's constant.

\section{BTZ solution}\label{se:2}
The Euclidean BTZ black hole \cite{Banados:1992wn,Banados:1992gq} has the topology of a solid torus, which we coordinatize by a radial coordinate $\rho\in[0,\infty)$, a contractible cycle coordinate $t\sim t+1$ and a non-contractible cycle coordinate $\phi\sim\phi+2\pi$. Since we work with fixed coordinate ranges the chemical potentials, essentially temperature and angular velocity, appear explicitly in the solutions.

The BTZ solution (in the basis of $sl(2,\mathbb{R})$ generators $L_+, L_0, L_-$) can be parametri\-zed by the dreibein $e = e_{\rho} d\rho + e_{\phi} d\phi + e_{t} dt$ as
\begin{align}
e_{\rho} & = \ell L_0 \qquad
e_{\phi}  =\frac{\ell}{2} \left(e^{\rho}(L_+ - L_-) - \frac{8\pi G}{\ell} e^{-\rho}(\mathcal{L}^+ L_- -\mathcal{L}^- L_+)  \right) \\
e_{t} & = \frac{\ell}{2} \left(e^{\rho}( \mu^+ L_+ + \mu^-L_-) - \frac{8\pi G}{\ell} e^{-\rho}(\mu^+\mathcal{L}^+ L_- + \mu^- \mathcal{L}^- L_+)  \right)\,.
\end{align}
Regularity of the solution requires a relation between charges $\mathcal{L}^\pm$ and chemical potentials $\mu^\pm$
\begin{equation}\label{regcon}
\mathcal{L}^\pm = \frac{\pi\ell}{8G(\mu^\pm)^2}
\end{equation}
where $\mu^{\pm}$ are related to the inverse temperature $\beta$ and the angular velocity $\Omega$ as
\begin{equation}\label{betamurel}
\beta = \frac{\ell}{2} (\mu^+ + \mu^-) \qquad \beta \Omega = - \frac12 (\mu^+ - \mu^-)\,.
\end{equation} 
In addition, the charges $\mathcal{L}^{\pm}$ are related to the mass $M$ and angular momentum $J$ as $\mathcal{L}^{\pm} = \frac{1}{4\pi}(M \ell \mp J)$ and to the loci of the BTZ Killing horizons $r_\pm$ as
\begin{equation}\label{BTZrpm}
r_\pm = \sqrt{8\pi G\ell \mathcal{L}^-} \pm \sqrt{8\pi G\ell \mathcal{L}^{+}} \,.
\end{equation}
The solution for $\omega$ follows from the constraint of vanishing torsion, which is one of the TMG equations of motion (EOM).
\begin{align}
\omega_{\rho} & = 0  \qquad \omega_{\phi}  =\frac{1}{2} \left(e^{\rho}(L_+ + L_-) - \frac{8\pi G}{\ell} e^{-\rho}(\mathcal{L}^+ L_- + \mathcal{L}^- L_+)  \right) \\
\omega_{t} & = \frac{1}{2} \left(e^{\rho}( \mu^+ L_+ - \mu^-L_-) - \frac{8\pi G}{\ell} e^{-\rho}(\mu^+\mathcal{L}^+ L_- - \mu^- \mathcal{L}^- L_+)  \right)
\end{align}
The auxiliary one-form $f$ also follows from the TMG EOM and is simply related to the dreibein $e$ by
\begin{equation}
f = -\frac{1}{2\ell^2 \mu}\, e\,.
\end{equation}

\section{On-shell action}\label{se:3}
Following Ba\~nados and Mendez \cite{Banados:1998ys} we compute the on-shell action using so-called `angular quantization'. The essence is to slice the solid torus into constant $\phi$ slices. Then on the constant $\phi$ slice we can use (any) regular coordinate system, while close to the boundary we use the $\rho, t, \phi$ Schwarzschild-like coordinates. The full action $\Gamma_{\scriptscriptstyle \mathrm{TMG}}=I_{\scriptscriptstyle \mathrm{TMG}}+B$ consists of the bulk action \eqref{STMG} and a boundary term $B$. In the angular decomposition for one-forms, 
$a = a_{\alpha} dx^{\alpha} + a_{\phi} d\phi$
where $a=e,\omega,f$ and
$x^{\alpha}$ are the coordinates on the disks of constant $\phi$, the full action is given by 
\begin{align} \label{angdecom}
\Gamma_{\scriptscriptstyle \mathrm{TMG}} = & \, \frac{1}{4\pi G} \int d\phi d^2 x \, \epsilon^{\alpha\beta} {\rm tr} \Big[ \sum_a a_{\phi} (\mathrm{EOM})_{\alpha\beta} 
- \sigma e_{\beta} \partial_{\phi} \omega_{\alpha} + f_{\beta} \partial_{\phi} e_{\alpha} - \frac{1}{2\mu} \omega_{\beta} \partial_{\phi} \omega_{\alpha} \Big] \nonumber \\
& + \frac{1}{4\pi G} \int_{\rho \to \infty} \!\!\!\!\!\!\!\!\!dtd\phi \,{\rm tr} \left[- \sigma e_t \omega_{\phi} + f_t e_{\phi} - \frac{1}{2\mu}\omega_t \omega_{\phi} \right]+ B\,.
\end{align}
The boundary term $B$ that gives a well-defined variational principle is one-half \cite{Mora:2004kb,Miskovic:2006tm,Detournay:2014fva} the Gibbons--Hawking--York boundary term. 
\begin{equation}
B = \frac{\sigma}{8 \pi G} \int_{\rho \to \infty}\!\!\!\!\!\!\!\!\! dtd\phi\, 
{\rm tr}(\omega_\phi e_t - \omega_t e_\phi)
\end{equation}

The bulk part of the action \eqref{angdecom} vanishes on-shell for spherically symmetric ($\phi$-independent) fields and we are left with the boundary terms in the second line of \eqref{angdecom}. Due to the compensating contribution from the boundary term $B$ the result is finite on the solutions given in the last section and it equals to $\beta F_{\textrm{\tiny BTZ}}$, where $F_{\textrm{\tiny BTZ}}$ is the BTZ free energy. Expressing everything in terms of temperature $T = \beta^{-1}$ and angular velocity $\Omega$ through \eqref{betamurel} gives the free energy of the BTZ black hole in TMG.
\begin{equation}
F_{\textrm{\tiny BTZ}} = \frac{1}{\beta} (I_{\scriptscriptstyle \mathrm{TMG}}^{\textrm{\tiny EOM}} + B) = - \frac{\pi^2 \ell^2 T^2}{2G(1-\Omega^2 \ell^2)} \left(\sigma + \mu^{-1}\Omega \right)
\label{eq:FBTZ}
\end{equation}
This result agrees with the free energy obtained by Legendre transforming entropy\footnote{%
We stress that for finite $\mu$ entropy \eqref{eq:SBTZ} does not obey the Bekenstein--Hawking area law. Nevertheless, it is compatible with the Cardy formula in the presence of a gravitational anomaly, i.e., the left- and right-moving central charges are not equal, $c-\bar c = 3/(\mu G)$ \cite{Kraus:2005zm}.
} \cite{Kraus:2005zm,Solodukhin:2005ah,Tachikawa:2006sz} 
\begin{equation}
 S_{\textrm{\tiny BTZ}} = -\frac{\partial F_{\textrm{\tiny BTZ}}}{\partial T}\Big|_{\Omega} = \sigma\,\frac{2\pi r_+}{4G} + \frac{2\pi r_-}{4G\mu\ell} 
 \label{eq:SBTZ}
\end{equation}
using the first law. We have thus succeeded in a first principles derivation of the BTZ free energy for TMG. 

Our derivation of the BTZ free energy \eqref{eq:FBTZ} from evaluating on-shell the full action \eqref{angdecom} readily generalizes to other cases. In the next section we focus particularly on asymptotically flat space, $\Lambda=0$, by considering the free energy of flat space cosmology solutions.

\section{Flat-space solutions}\label{se:4}
Consider locally flat line-elements with constant chemical potentials $\mu_{M}$, 
$\mu_{L}$ \cite{Gary:2014ppa}.
\begin{align} \label{FSC}
ds^2 = & \left(r^2 \mu_L^2 + \mathcal{M} (1+\mu_M)^2 + 2\mathcal{N}(1+\mu_M)\mu_L \right) du^2 \nonumber \\
& + \left(r^2 \mu_L + \mathcal{N} (1+\mu_M)\right) 2 du d\phi - (1+\mu_M) 2 dr du + r^2 d\phi^2
\end{align}
Analogously to the BTZ case discussed in the previous two sections, regularity of the flat space cosmology solutions \eqref{FSC} relates the charges $\mathcal{M}$, $\mathcal{N}$ to the chemical potentials $\mu_M$, $\mu_L$ and to the temperature $T$ and angular velocity $\Omega$. We find
\begin{equation} \label{FSCregcon}
\mathcal{M} = \frac{4\pi^2}{\mu_L^2}   \qquad \mathcal{N} = -\mathcal{M} \frac{1+\mu_M}{\mu_L} \qquad
T = \frac{1}{2\pi} \frac{\mathcal{M}^{3/2}}{|\mathcal{N}|}   \qquad \Omega = \frac{\mathcal{M}}{\mathcal{N}}\,.
\end{equation}
The easiest way to compute the free energy of these solutions in TMG is to write the dreibein which squares to \eqref{FSC} in the $sl(2,\mathbb{R})$ basis of the previous section and compute the corresponding spin-connection and auxiliary field from the TMG EOM. Then we can plug these solutions into the angularly decomposed action \eqref{angdecom} and eventually find the free energy.

Such a dreibein can be written as $e = e_r dr + e_\phi d\phi+ e_u du$ with
\begin{align}
e_r & = \frac12 L_-  \qquad 
e_{\phi} = - \frac{\mathcal{N}}{2} L_{-} + r L_0\\ 
e_u & = (1+\mu_M)L_+ - \Big(\frac14 \mathcal{M} (1+\mu_M) + \frac12 \mathcal{N} \mu_L \Big) L_- + r \mu_L L_0\,.
\end{align}
By solving the TMG EOM with this dreibein one finds that $f = 0$ and 
\begin{equation}
\omega_r = 0 \qquad \omega_{\phi}  = L_+ - \frac{\mathcal{M}}{4} L_-  \qquad 
\omega_u = \mu_L \Big( L_+ - \frac{\mathcal{M}}{4} L_- \Big)\,.
\end{equation}
After plugging this into the full action \eqref{angdecom} and using \eqref{FSCregcon} to write everything in terms of temperature and angular velocity we obtain the free energy.
\begin{equation}
F_{\textrm{\tiny FSC}} = \frac{1}{\beta} (I_{\scriptscriptstyle \mathrm{TMG}}^{\textrm{\tiny EOM}} + B) = - \frac{\pi^2 T^2}{2G \Omega^2} \left(\sigma + \frac{\Omega}{\mu} \right)
\end{equation}
This result agrees with the one derived in \cite{Bagchi:2013lma} by Legendre transforming the entropy of TMG with asymptotically flat boundary conditions. In particular, for flat space chiral gravity we obtain the entropy
\begin{equation}
S=-\frac{\partial F_{\textrm{\tiny FSC}}(\sigma=0)}{\partial T}\Big|_{\Omega} = \frac{\pi^2 T}{G\mu \Omega} = 2\pi \sqrt{\frac{ch}{6}} \label{eq:S}
\end{equation}
with $k=1/(8G\mu)>0$, $c=24k$ and $h=k\mathcal{M}$ \cite{Bagchi:2012yk}. The last equality in \eqref{eq:S} shows consistency with the chiral Cardy formula, as observed first in \cite{Bagchi:2013lma}, compatible with the flat space chiral gravity conjecture \cite{Bagchi:2012yk}.

\section{Correlation functions}\label{se:5}

Let us now move on from 0- to 1-point functions. They provide the first entries in the flat space holographic dictionary by identifying sources and vacuum expectation values as non-normalizable and normalizable solutions of the linearized EOM on the gravity side, respectively. The second order formulation \cite{Costa:2013vza,Fareghbal:2013ifa,Detournay:2014fva} reproduces the canonical results for the conserved charges in flat space Einstein gravity \cite{Barnich:2006av}. It is slightly easier to obtain these results in the first order formulation \cite{Bagchi:2015wna}. It would be interesting to generalize them to TMG in order to provide another check of the flat space chiral gravity conjecture. This would require either an extension from our calculations in the Chern--Simons formulation to Chern--Simons-like theories such as TMG or an application of the Horne--Witten formulation of CSG \cite{Horne:1988jf}.

If the 1-point functions match, as we expect them to do, one can actually go much further and check $n$-point correlation functions of the flat space holographic stress tensor in flat space chiral gravity. The procedure would follow the steps of the recent derivation of all $n$-point correlation functions in flat space Einstein gravity \cite{Bagchi:2015wna}, which we now summarize briefly:
\begin{itemize}
 \item Instead of directly calculating the $n$-th variation of the full action and inserting non-normalizable solutions to the linearized EOM we calculate the 1-point function on an arbitrary background, deformed by a chemical potential. On the field theory side this corresponds to a deformation of the original action $\Gamma_0$ to a deformed action $\Gamma_\mu$ with
 \begin{equation}
  \Gamma_\mu = \Gamma_0 - \int d^2z\, \mu(z,\bar z) {\cal O}(z,\bar z)
 \end{equation}
 where the chemical potential for the operator $\cal O$ is localized at $n-1$ points, $\mu = \sum_{i=2}^n \epsilon_i \delta(z-z_i,\bar z -\bar z_i)$; the coefficients $\epsilon_i$ are a convenient book keeping device and $z,\bar z$ are some coordinates used in the 2-dimensional field theory.
  \item The 1-point function on the deformed background then yields the $n$-point function for the original background, e.g.~for $n=2$ we get
  \begin{equation}
   \langle{\cal O}(z_1,\bar z_1)\rangle_\mu = \langle{\cal O}(z_1,\bar z_1)\rangle_0 + \epsilon_2 \langle {\cal O}(z_1,\bar z_1){\cal O}(z_2,\bar z_2)\rangle_0 + \dots
  \end{equation}
  The term to linear order in $\epsilon_2$ yields the 2-point correlator, both on gravity and field theory sides. The same procedure works for arbitrary $n$-point correlators.
  \item In order to show the equivalence of all correlations functions on gravity and field theory sides one can use complete induction by proving recursion relations between $n$- and $(n-1)$-point correlation functions, analogous to the BPZ-recursion relations for the stress tensor in a CFT \cite{Belavin:1984vu}. For flat space Einstein gravity and Galilean CFTs these recursion relations were established recently \cite{Bagchi:2015wna}, thus showing the equivalence of all flat space holographic stress tensor correlation functions with corresponding Galilean CFT correlation functions.
\end{itemize}
This procedure provides a fairly non-trivial check of flat space holography in three dimensions. It would be great to generalize it to flat space chiral gravity.
 
\section{Flat space holographic entanglement entropy}\label{se:6}

As concluding part of this proceedings contribution we focus on a further check of flat space holography. One particularly interesting part of the AdS/CFT developments was the insight by Ryu and Takayanagi a decade ago that entanglement entropy can be calculated by elementary methods on the gravity side, through minimizing the area of certain hypersurfaces, depending on the entangling region for which entanglement entropy is calculated \cite{Ryu:2006bv}. With methods similar to the ones used in the CFT derivation \cite{Holzhey:1994we,Vidal:2002rm,Calabrese:2004eu} one can also derive entanglement entropy for 2-dimensional Galilean CFTs \cite{Bagchi:2014iea} and, following the holographic computation of entanglement entropy in the presence of a gravitational Chern--Simons term \cite{Castro:2014tta}, we expect that this result should match with the flat space chiral gravity prediction
\begin{equation}
 S_{\textrm{EE}} = \frac{c}{6}\,\ln{\frac{L}{a}}
\end{equation}
where $c=24k$, $L$ is the length of the entangling region and $a$ is an ultraviolet cutoff. Also this prediction of flat space chiral gravity was confirmed recently \cite{Hosseini:2015uba}.

\section*{Acknowledgments}

We thank Arjun Bagchi for collaboration on $n$-point correlation functions in flat space holography. DG additionally thanks Arjun Bagchi, Stephane Detournay, Max Riegler, Jan Rosseel and Joan Simon for a wonderful long-term collaboration on numerous aspects of flat space holography.

DG was supported by projects of the Austrian Science Fund (FWF) Y~435-N16, I~952-N16 and I~1030-N27, and by the program
Science without Borders, project CNPq-401180/2014-0.
WM was supported by the FWF project P~27182-N27.


\providecommand{\href}[2]{#2}\begingroup\raggedright\endgroup

\end{document}